\documentclass[twocolumn,superscriptaddress,aps,pra]{revtex4-2}

\makeatletter
\newcommand*{\rom}[1]{\expandafter\@slowromancap\romannumeral #1@}
\makeatother

\usepackage{color}
\usepackage{graphicx}
\usepackage{hyperref}
\usepackage{amsmath,amssymb}
\usepackage{epstopdf}
\usepackage{subcaption}
\usepackage{float}

\usepackage{ragged2e} 
\graphicspath{{figs/}}

\def\beq{\begin{equation}}
\def\eeq{\end{equation}}
\def\bea{\begin{eqnarray}}
\def\eea{\end{eqnarray}}

\usepackage{xcolor}

\begin{document}
\pagestyle{plain}

\title{Thermodynamic geometry of inclusion statistics}
\author{M. H. Naghizadeh Ardabili}
\affiliation{Department of Physics, University of Mohaghegh Ardabili, P.O. Box 179, Ardabil, Iran}
\author{Omid Yahyayi Monem}
\affiliation{Department of Physics, University of Mohaghegh Ardabili, P.O. Box 179, Ardabil, Iran}
\author{Habib Esmaili}
\affiliation{Department of Physics, University of Mohaghegh Ardabili, P.O. Box 179, Ardabil, Iran}
\author{Zahra Ebadi}
\affiliation{Department of Physics, University of Mohaghegh Ardabili, P.O. Box 179, Ardabil, Iran}
\author{Hosein Mohammadzadeh}
\email{mohammadzadeh@uma.ac.ir}
\affiliation{Department of Physics, University of Mohaghegh Ardabili, P.O. Box 179, Ardabil, Iran}

\begin{abstract}
We investigate the thermodynamic geometry of an ideal quantum gas obeying
inclusion statistics, characterized by a negative statistical parameter $g < 0$. In this framework the grand-canonical partition function admits a finite maximum fugacity, and the thermodynamic scalar curvature $R$ is strictly positive for all $g < 0$, reflecting effective attractive statistical interactions analogous to those of a bosonic system. As the fugacity approaches its maximum value, $R$ diverges, signaling a phase transition of the Bose-Einstein condensation type. A key distinction from the ordinary ideal Bose gas is that the condensation temperature is elevated relative to the bosonic case, and finite-temperature condensation occurs even in the dimensional regime $1/2 < D/\sigma \leq 1$ where standard bosons do not condense, while for $D/\sigma \leq 1/2$ the transition temperature vanishes. Three independent criteria; divergence of $R$, the maximum fugacity singularity, and the non-analytic cusp in the specific heat, coincide at the same condensation point, confirming the thermodynamic consistency of the transition.
\end{abstract}

\maketitle
\section{Introduction}
Quantum statistics governs the collective behavior of many-particle systems and underlies some of the most striking phenomena in statistical and condensed-matter physics. Identical particles are classified according to the exchange symmetry of their wave function: bosons obey Bose-Einstein (BE) statistics and can occupy a single quantum state without restriction, whereas fermions obey Fermi-Dirac (FD) statistics and are forbidden from multiple occupancy by the Pauli exclusion principle \cite{pathria2011statistical}. The most celebrated consequence of BE statistics is Bose--Einstein condensation (BEC), the macroscopic accumulation of particles in the single-particle ground state below a critical temperature. A well-known feature of the ideal Bose gas is that, in the absence of a confining potential, finite-temperature condensation occurs only for spatial dimension $D\geq3$; in lower dimensions thermal fluctuations prevent the formation of a macroscopically occupied ground state \cite{huang2008statistical}.
 
Real particles need not, however, be confined to these two limiting cases. The possibility of statistics interpolating between bosons and fermions has motivated the development of generalized statistical frameworks \cite{haldane1991fractional,wu1995statistical,polychronakos1996probabilities,yahyayi2026thermodynamic,yan2021statistical,chung2017duality}. Among these, Haldane's fractional exclusion statistics (FES) offers a purely combinatorial generalization, in which a statistical parameter $g$ measures the reduction of the accessible single-particle Hilbert space caused by the addition of a further particle \cite{haldane1991fractional}. The bosonic and fermionic limits are recovered for $g=0$ and $g=1$, respectively, while $0<g<1$ describes intermediate exclusion. The corresponding equilibrium occupation-number distribution was derived by Wu \cite{wu1995statistical}, and the thermodynamic geometry of an ideal FES gas in arbitrary spatial dimension was subsequently obtained in closed form \cite{mirza2010thermodynamic}. Within this framework, the thermodynamic properties interpolate continuously between the Bose and Fermi limits, but BEC survives only in the strict bosonic point $g=0$: any finite exclusion, however weak, is sufficient to destroy condensation.
 
More recently, Ouvry and Polychronakos proposed a complementary extension of Haldane's combinatorial construction to negative values of the statistical parameter, $g<0$, obtained by analytic continuation of the underlying counting formula \cite{ouvry2023inclusion}. Rather than exclusion, particles governed by this extension display statistical inclusion: the occupation of an already-populated state is enhanced relative to ordinary bosons. This is not merely a formal extrapolation. Inclusion statistics predicts a finite maximum fugacity that depends explicitly on the statistical parameter, in place of the bosonic bound $z_{\max}=1$, and, most notably, it lowers the critical dimension for condensation from $D=3$ to $D=2$: an ideal $g$-inclusion gas can condense at finite temperature already in two dimensions, with the condensation temperature increasing monotonically as the inclusion becomes stronger, i.e., as $g$ becomes more negative.
 
While the statistical and thermodynamic foundations of inclusion statistics have thus been established, its critical behavior has so far been characterized only through the occupation-number distribution and the associated fugacity bounds, and a detailed account of its condensation mechanism and thermodynamic signatures is still lacking. A complementary and physically transparent viewpoint is offered by thermodynamic geometry, in which the space of thermodynamic parameters is endowed with a Riemannian metric most naturally the Fisher-Rao metric obtained from the second derivatives of the logarithm of the partition function so that thermodynamic fluctuations acquire a geometric interpretation \cite{weinhold1975metric,ruppeiner1979thermodynamics,ruppeiner1995riemannian}. The associated scalar curvature encodes the effective statistical interactions among particles and is known to diverge at continuous phase transitions, providing a diagnostic of criticality that is independent of, yet consistent with, the standard thermodynamic treatment; for the ideal Bose gas, for example, the curvature diverges precisely at $z=1$ \cite{janyszek1990riemannian}. For FES with $g>0$ this geometric structure varies smoothly with $g$ and becomes singular only at the bosonic point $g=0$, where BEC occurs. This geometric perspective has been successfully employed to investigate a broad range of classical, quantum, and generalized statistical systems \cite{esmaili2024thermodynamic,seifi2025intrinsic,mohammadzadeh2016perturbative,mirza2011condensation,guzman2018geometric, quevedo2022geometrothermodynamics,zaldivar2023ideal,seifi2025mittag,ardabili2025haldane,mohammadzadeh2026thermodynamic,esmaili2024thermodynamic,seifi2025intrinsic,mohammadzadeh2016perturbative,mirza2011condensation}
 
In this work, we extend the thermodynamic-geometry approach to the hitherto unexplored regime of negative statistical parameter and present a geometric characterization of condensation in ideal $g$-inclusion statistics. We show that, in marked contrast to ordinary exclusion statistics, the thermodynamic curvature diverges at a finite critical fugacity $z_c(g)<1$ for every $g<0$, and that this geometric critical point coincides both with the threshold fugacity obtained from the occupation-number distribution and with the maximum fugacity predicted by the statistical formulation, thereby unifying three independent characterizations of the same transition within a single geometric picture. Building on the resulting critical fugacity, we obtain the condensation temperature and examine its dependence on the statistical parameter $g$ and on the effective dimensionality $D/\sigma$ of the system, confirming that finite-temperature condensation persists down to two dimensions and is enhanced by stronger statistical inclusion. Finally, we analyze the specific heat at constant volume across the transition and show that it develops the non-analytic behavior characteristic of a continuous phase transition.
 
The remainder of this paper is organized as follows. In Sec.\ref{A Review of Haldane Exclusion Statistics}, we briefly review Haldane fractional exclusion statistics and its thermodynamics. In Sec.\ref{Inclusion statistics}, we introduce inclusion statistics and its associated maximum fugacity. In Sec.\ref{Thermodynamic Geometry and Critical Behavior of Inclusion Statistics}, we apply thermodynamic geometry to inclusion statistics, determine the critical fugacity and condensation temperature, and examine the specific heat across the transition. Section\ref{CONCLUSIONS} summarizes our main conclusions.

\section{A Review of Haldane Exclusion Statistics}\label{A Review of Haldane Exclusion Statistics}

In conventional quantum statistics, particles are classified into two limiting categories according to the way they occupy single-particle quantum states. Bosons obey BE statistics, allowing an arbitrary number of particles to occupy the same quantum state, whereas fermions obey FD statistics and satisfy the Pauli exclusion principle, which restricts each quantum state to at most one particle.

Haldane introduced a generalized statistical framework known as 
fractional exclusion statistics (FES), in which a statistical parameter 
$g$ continuously interpolates between BE and FD statistics \cite{haldane1991fractional}. 
In this framework, the value of the statistical parameter determines the strength 
of statistical exclusion between the bosonic and fermionic limits. The central idea of Haldane statistics is that particle statistics are characterized by the variation of the dimension of the accessible single-particle Hilbert space as particles are added to the system. For a single species of indistinguishable particles, this relation is expressed as:
\begin{equation}
g = -\frac{d_{N+\Delta N} - d_N}{\Delta N},
\end{equation}
where $\Delta d=d_{N+\Delta N} - d_N$ denotes the change of the dimension of the accessible single-particle Hilbert space
caused by adding $\Delta N$ particles, and $g$ is the statistical
exclusion parameter. This relation provides a simple physical interpretation of the statistical parameter. For bosons ($g=0$), the number of available states remains unchanged as particles are added, reflecting the absence of statistical exclusion. For fermions ($g=1$), each additional particle removes one available state, recovering the Pauli exclusion principle. Intermediate values $0<g<1$ correspond to fractional exclusion, where each added particle decreases the number of accessible states by a fraction $g$. In this regime, the exclusion effect is
stronger than that of bosons but weaker than the complete Pauli exclusion obeyed by fermions.

Using the generalized exclusion principle together with the maximum entropy formalism, Wu derived the equilibrium distribution function for particles obeying fractional exclusion statistics \cite{wu1995statistical}. The average occupation number $n_i$ satisfies the implicit relation:
\begin{equation}\label{DF}
(1-g n_i)^g
\left[1+(1-g)n_i\right]^{1-g}
=
n_i\, e^{(\epsilon_i-\mu)/(k_B T)}
\end{equation}
where $\epsilon_i$ is the single-particle energy, $\mu$ is the chemical potential, $T$ is the temperature, and $k_B$ is the Boltzmann constant. Introducing the fugacity $z=e^{\beta\mu}$, with $\beta=(k_B T)^{-1}$, the occupation number can be written as $n_i=n(g,\beta,z,\epsilon_i)$, emphasizing that the statistical distribution depends on the exclusion parameter $g$, the temperature, the fugacity, and the single-particle energy.

In the thermodynamic limit, an ideal gas obeying fractional exclusion statistics in a $D$-dimensional space has been investigated~\cite{mirza2010thermodynamic}. By introducing the auxiliary variable $w$ through the transformation from $\epsilon$ to $w$,
\begin{equation}\label{BEMu}
\beta(\epsilon-\mu)=g\ln w+(1-g)\ln(1+w),
\end{equation}
with the corresponding differential transformation,
\begin{equation}
d\epsilon=\frac{1}{\beta}\frac{w+g}{w(1+w)}dw,
\end{equation}
the thermodynamic quantities, including the total particle number and the total internal energy, can be expressed as:
\begin{align}
N = \int_0^\infty n(\epsilon)\, \Omega(\epsilon) \, d\epsilon = \frac{A^D}{\Gamma(D/\sigma)} \beta^{-D/\sigma} \mathcal{H}_{D/\sigma}(g,z),
\label{eq:TN}
\\
U = \int_0^\infty \epsilon\, n(\epsilon)\, \Omega(\epsilon) \, d\epsilon = \frac{A^D}{\Gamma(D/\sigma)} \beta^{-D/\sigma-1} \mathcal{H}_{D/\sigma+1}(g,z).
\label{eq:U}
\end{align} 
where ${H}_{\nu}(g,z)$ is:
\begin{equation}\label{HD}
\mathcal{H}_{\nu}(g,z)=
\int_{w_0}^{\infty}
\frac{
\left[
\ln\left(z w^{g}(1+w)^{1-g}\right)
\right]^{\nu-1}
}
{w(1+w)}
\,dw ,
\end{equation}
and $A=L\sqrt{\pi}/\left(a^{1/\sigma}h\right)$ is a constant, $\sigma$ is the dispersion exponent, $L$ is the length of the box, $h$ is Planck's constant, $a$ is a proportionality constant, $\Gamma(x)$ denotes the Gamma function, and the lower integration limit $w_0$ is obtained from Eq.~\eqref{BEMu} by setting $\epsilon=0$.
    


\section{Inclusion statistics}\label{Inclusion statistics} 

The thermodynamics of inclusion statistics can be obtained by analytically 
continuing the combinatorial formulation of Haldane fractional exclusion 
statistics from the conventional regime $g>0$ to negative values of the 
statistical parameter, $g<0$ \cite{ouvry2023inclusion}. In this framework, the grand partition function 
for a system containing $K$ single-particle states with energy $\epsilon$ is 
defined as
\begin{equation}
\mathcal{Z}(K,z)
=
\sum_{N=0}^{\infty}
G_g(K,N)z^N ,
\end{equation}
where $G_g(K,N)$ denotes the number of available many-particle states.
In the 
thermodynamic limit, $K\gg1$, the grand partition function becomes extensive 
and can be written as
\begin{equation}
\ln \mathcal{Z}=K\ln y+O(K^{-1}),
\end{equation}
where $y$ can be interpreted as the effective grand partition function 
associated with a single-particle energy level. The function $y(z,g)$ satisfies 
the implicit relation
\begin{equation}
y^g-y^{g-1}=z .
\end{equation}
For negative values of the statistical parameter, this equation is 
understood through analytic continuation of the exclusion statistics framework. 
The physical solution corresponds to the branch satisfying $y>1$. An important 
feature of inclusion statistics is the existence of a maximum allowed fugacity. 
Unlike the Bose case, where the critical value is $z_{\max}=1$, the maximum 
fugacity in inclusion statistics depends explicitly on the statistical parameter 
and is obtained from the extremum condition
\begin{equation}
\frac{\partial z}{\partial y}=0 .
\end{equation}
which determines the maximum value of the function $z(y,g)$. 

\begin{equation}
y_{\max}=\frac{g-1}{g},
\end{equation}
and consequently the maximum fugacity for $g<0$ is obtained as:
\begin{equation}\label{zmax}
z_{\max}=(-g)^{-g}(1-g)^{(g-1)}.
\end{equation}

\begin{figure}[htbp]
    \centering
    \includegraphics[width=1\linewidth]{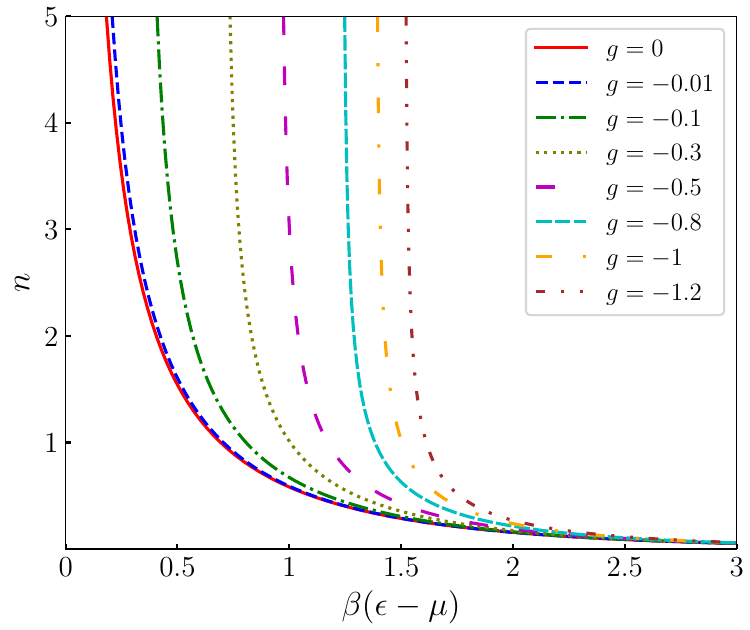}
    \caption{\justifying
    	Distribution function of \(g\)-inclusion statistics as a function of 
    \(\beta(\epsilon-\mu)\) for different values of \(g\).}
    \label{fig:inclusion_distribution}
\end{figure}

We further examine the behavior of the distribution function as a function of 
$\beta(\epsilon-\mu)$ for negative values of the statistical parameter $g$. 
Using the distribution relation given in Eq.~\ref{DF}, the corresponding curves are 
obtained and presented in Fig.~\ref{fig:inclusion_distribution}. It can be observed that for $g<0$, the 
distribution exhibits an enhanced occupation behavior compared with the 
conventional BE statistics, reflecting the inclusion nature of the 
statistics. It is also evident that the distribution function diverges for each value of
the statistical parameter $g$, indicating the existence of a threshold point
$x_{th}=\beta(\epsilon-\mu)$. By setting $\epsilon=0$, corresponding to the
ground state, one obtains $x_{th}=-\beta\mu$. Using the relation between the
chemical potential and the fugacity, the corresponding threshold fugacity can
therefore be determined. This result will be discussed in the following
section.

\section{Thermodynamic Geometry and condensation Behavior of Inclusion Statistics}\label{Thermodynamic Geometry and Critical Behavior of Inclusion Statistics}

In this section, we employ the framework of thermodynamic geometry to investigate the thermodynamic structure and behavior of inclusion statistics. In this approach, the space of thermodynamic parameters is regarded as a Riemannian manifold. Originally introduced by Weinhold and Ruppeiner, thermodynamic geometry establishes a connection between thermodynamic fluctuations and geometric quantities, with the scalar curvature encoding information about the underlying statistical interactions of the system.\cite{ruppeiner1979thermodynamics,weinhold1975metric}
Within the framework of thermodynamic geometry, divergences of the thermodynamic curvature are often associated with phase transitions and provide valuable information about critical phenomena, such as condensation points and the corresponding critical values of the fugacity or chemical potential. For instance, in an ideal Bose gas, the divergence of the thermodynamic curvature at $z=1$ signals the onset of BE condensation and identifies the critical point of the system.
 
Among the various formulations of thermodynamic geometry, the Fisher--Rao information metric provides a particularly suitable description for quantum statistical systems. This metric is obtained from the second derivatives of the logarithm of the partition function with respect to the non-extensive thermodynamic parameters and is defined as
\begin{equation}
G_{ij}=\partial_i\partial_j \ln \mathcal{Z}.
\end{equation}
For a two-dimensional thermodynamic parameter space, the scalar curvature
$R$ can be expressed directly in terms of the metric components and their
derivatives. For Haldane fractional exclusion statistics, the metric
components and their derivatives were previously calculated in
Eqs.~(18) and (22) of Ref.~\cite{mirza2010thermodynamic}. It has been shown that the thermodynamic curvature exhibits an intermediate behavior between the two limiting quantum statistics. In particular, the cases
$g=0$ and $g=1$ recover the thermodynamic geometry of ideal BE
and FD gases, respectively, while intermediate values of the
statistical parameter describe a continuous interpolation between these two
limits. It should be noted that, except for the case $g=0$, where Bose
condensation occurs, no condensation is observed for other values of the
statistical parameter.

\begin{figure}[!htbp]
    \centering
    
    \begin{subfigure}{0.5\textwidth}
        \centering
        \includegraphics[width=\textwidth]{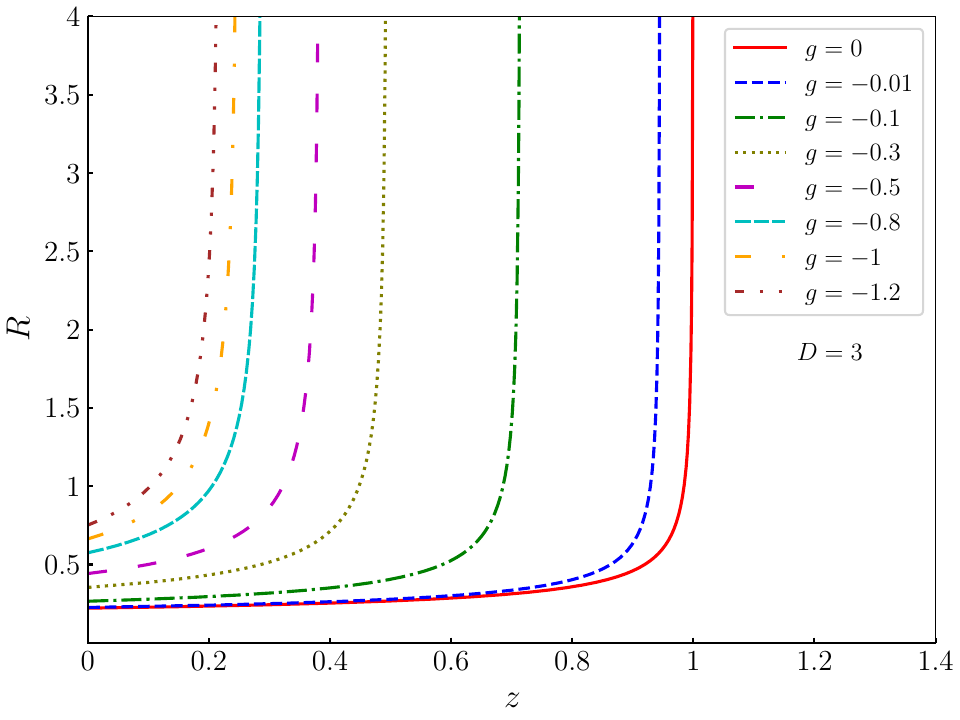}
        \caption{}
        \label{fig:curvature_3D}
    \end{subfigure}
    \hfill
    \begin{subfigure}{0.5\textwidth}
        \centering
        \includegraphics[width=\textwidth]{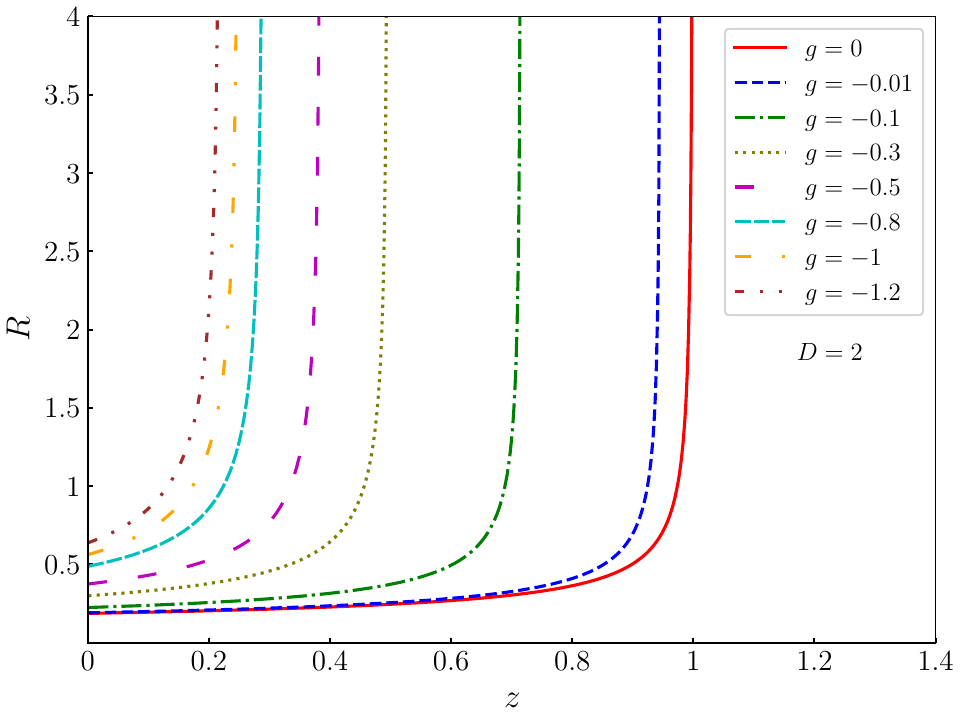}
        \caption{}
        \label{fig:curvature_2D}
    \end{subfigure}

    \caption{\justifying
    	Thermodynamic curvature as a function of fugacity for isothermal processes 
    (\(\beta=1\)). (a) In a 3-dimensional ideal \(g\)-inclusion-statistics gas with 
    \(\sigma=2\) (non-relativistic regime). (b) In a 2-dimensional ideal 
    \(g\)-inclusion-statistics gas with \(\sigma=1\) (ultra-relativistic regime).}
    
    \label{fig:thermodynamic_curvature}
\end{figure}

In the present work, we extend this analysis to negative values of the
statistical parameter $g$. The thermodynamic curvature as a function of the
fugacity is shown in Fig.~\ref{fig:thermodynamic_curvature} for the cases
$D=3$ with $\sigma=2$ and $D=2$ with $\sigma=1$ for different values of
$g$. As expected, the bosonic limit corresponding to $g=0$ is correctly
recovered. In contrast, for negative values of $g$, which characterize
inclusion statistics, the thermodynamic curvature exhibits a divergence at
a critical fugacity for each value of $g<0$. This critical fugacity is
always smaller than the bosonic critical value, $z=1$, and decreases
monotonically as $g$ becomes more negative.

\begin{figure}[!htbp]
    \centering
    \includegraphics[width=1\linewidth]{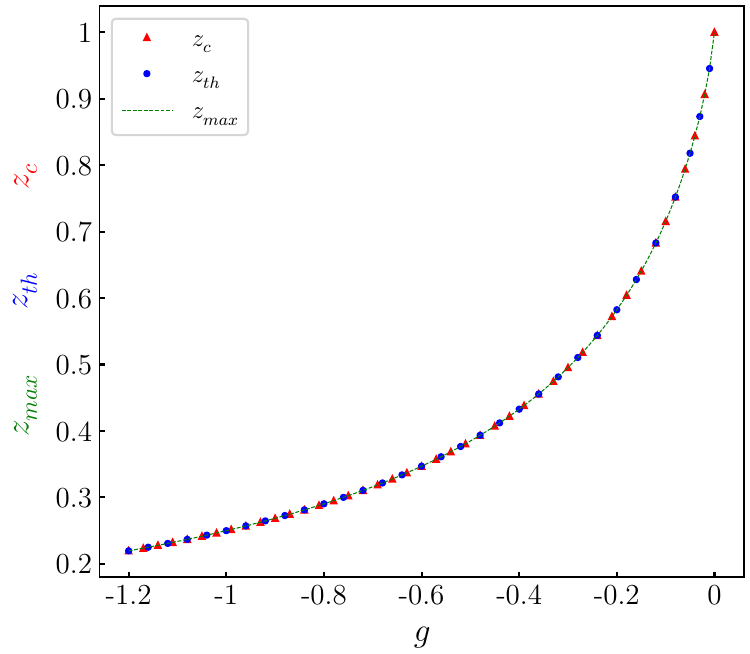}
    \caption{\justifying
    	critical fugacity $z_c$, threshold fugacity $z_{th}$, and maximum fugacity $z_{\max}$ as functions of the statistical parameter $g$ for $D=3$ and $\sigma=2$.}
    \label{fig5}
\end{figure}

Figure~\ref{fig5} presents the dependence of the critical fugacity $z_c$ on the
statistical parameter $g$ in the negative-$g$ regime, for $D=3$ and $\sigma=2$.
The critical fugacities, identified by the divergence of the thermodynamic
curvature in Fig.~\ref{fig:thermodynamic_curvature}, are compared with the
threshold fugacities $z_{th}$ extracted from the distribution function in
Fig.~\ref{fig:inclusion_distribution}, and with the maximum fugacities $z_{\max}$
obtained from Eq.~\eqref{zmax}. The three independent criteria yield mutually
consistent estimates of the condensation threshold across the entire range of
$g$, establishing that the curvature divergence, the onset of the
inclusion-distribution singularity, and the fugacity bound $z_{\max}$ all
coincide with the same critical point of the inclusion statistics.

We now investigate the behavior of the thermodynamic curvature as a function of 
the reduced condensation temperature. For a fixed particle density $N/V$, the 
transition temperature is determined from the particle-number equation. Therefore, the particle density can be 
expressed as
\begin{equation}\label{NVTZ}
\frac{N}{V}
=
\frac{1}{\Gamma(D/\sigma)}
\left(
\frac{\sqrt{\pi}}{a^{1/\sigma}h}
\right)^D
\beta^{-D/\sigma}
\mathcal{H}_{D/\sigma}(g,z).
\end{equation}

\begin{figure}[t]
	\centering
	
	\begin{subfigure}[b]{0.49\columnwidth}
		\centering
		\includegraphics[width=\textwidth]{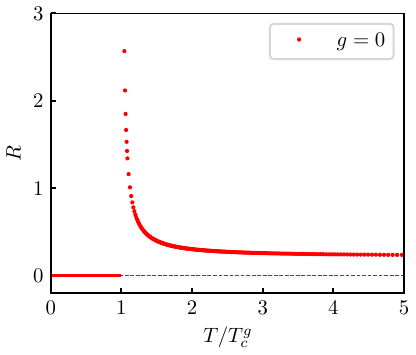}
		\caption{}
		\label{fig:a}
	\end{subfigure}
	\hfill
	\begin{subfigure}[b]{0.49\columnwidth}
		\centering
		\includegraphics[width=\textwidth]{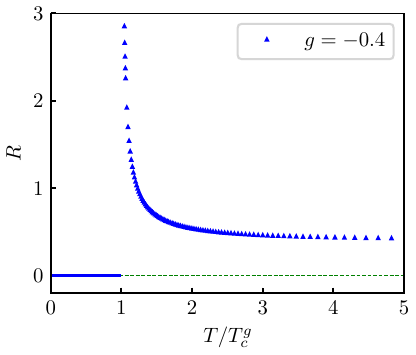}
		\caption{}
		\label{fig:b}
	\end{subfigure}
	
	\vspace{0.15cm}
	
	\begin{subfigure}[b]{0.49\columnwidth}
		\centering
		\includegraphics[width=\textwidth]{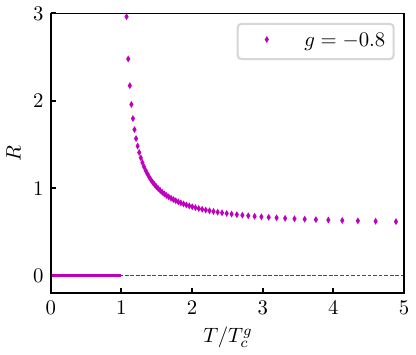}
		\caption{}
		\label{fig:c}
	\end{subfigure}
	\hfill
	\begin{subfigure}[b]{0.49\columnwidth}
		\centering
		\includegraphics[width=\textwidth]{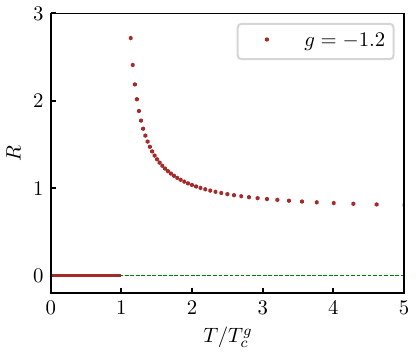}
		\caption{}
		\label{fig:d}
	\end{subfigure}
	
	\caption{\justifying Thermodynamic curvature as a function of the reduced temperature $T/T_c^{g}$  in a three-dimensional ideal $g$-inclusion-statistics gas with $\sigma=2$. The curves are shown for $g=0,\,-0.4,\,-0.8,\,-1.2$. The divergence of the curvature at $T/T_c^{g}=1$ indicates the presence of a phase transition and condensation.}
	\label{fig:thermodynamic_curvature_TTc}
\end{figure}

For comparison, the BE condensation temperature is obtained in the 
limit $g=0$ and $z_c=1$, yielding
\begin{equation}
k_B T_c^{BE}
=
\frac{h^2}{2m\pi}
\left[
\frac{N}{V\,\zeta\left(D/\sigma\right)}
\right]^{\sigma/D}.
\end{equation}
Therefore, the ratio between the $g$-inclusion transition temperature and the
conventional BE condensation temperature is given by
\begin{equation}\label{TTC}
T_c^{g}
=
\left[
\frac{
\zeta(D/\sigma)
}
{
\mathcal{H}_{D/\sigma}(g,z_c(g))
}
\right]^{\sigma/D}T_c^{BE},
\end{equation}
which explicitly demonstrates the enhancement of the condensation temperature
in the negative-$g$ inclusion regime.
For a fixed value of $N/V$, we obtain the fugacity as a function of temperature numerically  using Eq.~\ref{NVTZ} for different
values of the statistical parameter $g$. Then, the thermodynamic curvature   is evaluated as a function of the reduced temperature; $T/T_c^{g}$is evaluated using Eq.~\ref{TTC}. The thermodynamic curvature for different values of inclusion parameter $g$ is shown in
Fig.~\ref{fig:thermodynamic_curvature_TTc}. The curves exhibit a
divergence at the critical point $T/T_c^{g}=1^{+}$, indicating the
occurrence of a condensation phase transition. Also, for condensation phase with $T<T_c^{g}$, the fugacity is fixed at $z=z_{max}$ and the thermodynamic fluctuating parameters space is one dimensional and trivially it is flat. Hence,
for $T/T_c^{g}<1$, the thermodynamic curvature $R$ vanishes.

%
%
%
\begin{figure}[t]
    \centering
    \includegraphics[width=1\linewidth]{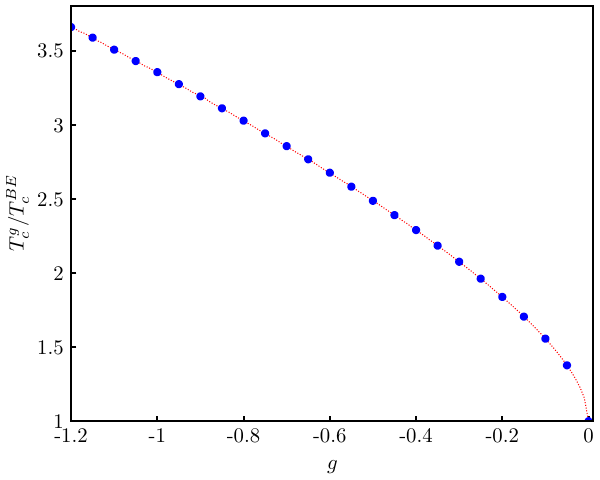}
    \caption{\justifying
    	Ratio of the g-inclusion condensation temperature to the ordinary BE condensation temperature, \(T_c^{g}/T_c^{BE}\), as a function of the Haldane statistical parameter \(g\), for \(D=3\) and \(\sigma=2\).}
    \label{figTTcg}
\end{figure}

\begin{figure}[t]
    \centering
    \includegraphics[width=1\linewidth]{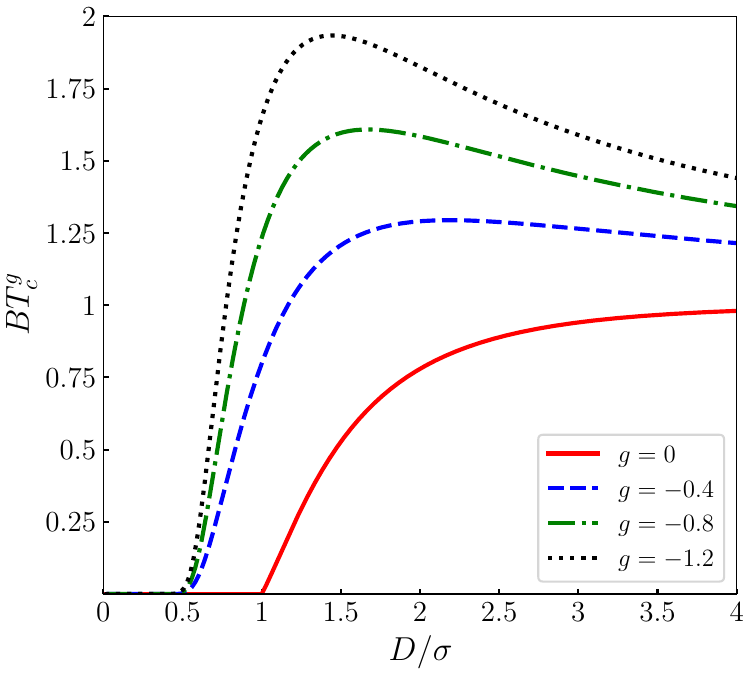}
    \caption{\justifying
    	Dimensionless condensation temperature 
    $\beta_c T_c^{\mathrm{g}}$ as a function of the effective dimensionality 
    $D/\sigma$ in the $g$-inclusion-statistics framework.}
    \label{fig:Tc_dimension}
\end{figure}

Having established the existence of condensation at a critical fugacity
$z_c(g)<1$, we now turn to the corresponding condensation temperature. Using Eq.~\ref{TTC}, the behavior of the ratio 
$T_c^{g}/T_c^{BE}$ as a function of the statistical 
parameter $g$ can be investigated. The corresponding results are presented in 
Fig.~\ref{figTTcg}. In the bosonic limit $g=0$, the standard result is recovered,
$T_c^{g}=T_c^{BE},$ whereas for negative values of $g$, the ratio
${T_c^{g}}/{T_c^{BE}}$
becomes larger than unity and increases monotonically as $g$ decreases. 
Motivated by the prediction that the lower critical dimension for condensation
is reduced from three to two dimensions in inclusion statistics, we further
investigate the dependence of the condensation temperature on the effective
dimensionality of the system. For this purpose, we define the quantity $B T_c^{g},$ where $B={2\pi m k_B}/{h^2}$ is a constant. 
The behavior of this quantity as a function of the effective dimensionality is
shown in Fig.~\ref{fig:Tc_dimension}. This analysis illustrates how the
critical temperature changes with the dimensionality of the system. It is observed that condensation occurs already
in two dimensions, whereas no condensation is found for effective
dimensionalities below two for ordinary bosons. In fact for $g=0$, the transition temperature for $D/\sigma\le 1$ tends to zero and the finite temperature condensation does not exist while for $g<0$, for $D/\sigma>1/2$, the condensation at finite temperature can occur. 

Now, we consider the two-dimensional case, where condensation becomes possible within the framework of $g$-inclusion statistics. For $D/\sigma=1$ (corresponding to $D=2$ and $\sigma=2$), by combining Eqs. \ref{BEMu}, \ref{zmax}, \ref{HD}, \ref{NVTZ} and noting that the condensation temperature in two dimensions depends solely on the statistical parameter $g$, the condensation temperature is obtained as:
\begin{equation}
T_c^{g}
=
\frac{n h^2}
{2\pi m k_B
\ln\!\left(\frac{1 - g}{-g}\right)}.
\label{Tc2D}
\end{equation}

At $g=0$, the condensation temperature vanishes, $T_c=0$. As $g$ decreases within the negative ($g$-inclusion) regime, the logarithmic denominator decreases monotonically, resulting in a corresponding monotonic increase of the condensation temperature.


Finally, we investigate the behavior of the specific heat at constant volume as 
a function of the reduced temperature $T/T_c^{g}$ for 
$D=3$ and $\sigma=2$ for different values of the inclusion parameter $g$. 
Near a phase transition, thermodynamic quantities may exhibit non-analytic 
features that reflect the critical behavior of the system. In particular, the 
specific heat of an ideal Bose gas develops a non-differentiable behavior at 
the condensation temperature.

Using Eq.~\ref{HD}, the heat capacity can be expressed analytically in both 
temperature regimes: below the transition point 
$T<T_c^{g}$ with fixed value of fugacity $z=z_{c}$ and above the critical temperature. 

Below the transition temperature,
$T<T_c^{g},$ the fugacity remains fixed at its critical value,
$z=z_c(g),$ and differentiating the internal energy with respect to temperature gives
\begin{equation}
\frac{C_V}{Nk_B}
=
\left(\frac{D}{\sigma}+1\right)
\frac{\mathcal{H}_{D/\sigma+1}(g,z_c)}
{\mathcal{H}_{D/\sigma}(g,z_c)}
\left(
\frac{T}{T_c^{g}}
\right)^{D/\sigma}.
\end{equation}

Above the critical temperature, the constant-volume heat capacity, $C_V = (\partial U/\partial T)_{V,N}$, can be evaluated by applying the chain rule over the intensive coordinates $(\beta, z)$ under the constraint $dN = 0$, which yields $(\partial z/\partial \beta)_N = -(\partial N/\partial \beta)_z / (\partial N/\partial z)_\beta$. Consequently, using Eqs. (18) in Ref. (\cite{mirza2010thermodynamic}), $C_V$ is directly mapped to the metric components as
\begin{equation}
C_V = k_B \beta^2 \left( G_{\beta\beta} - \frac{G_{\beta\gamma}^2}{G_{\gamma\gamma}} \right).
\end{equation}

\begin{figure}[t]
	\centering
	
	\begin{subfigure}[b]{0.49\columnwidth}
		\centering
		\includegraphics[width=\textwidth]{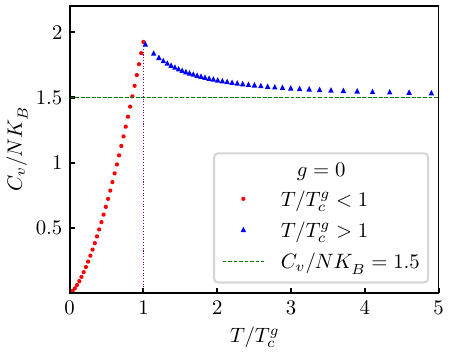}
		\caption{}
		\label{fig:a}
	\end{subfigure}
	\hfill
	\begin{subfigure}[b]{0.49\columnwidth}
		\centering
		\includegraphics[width=\textwidth]{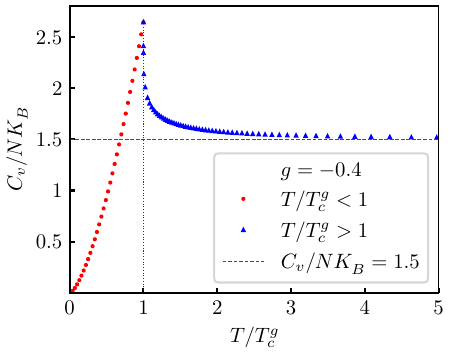}
		\caption{}
		\label{fig:b}
	\end{subfigure}
	
	\vspace{0.15cm}
	
	\begin{subfigure}[b]{0.49\columnwidth}
		\centering
		\includegraphics[width=\textwidth]{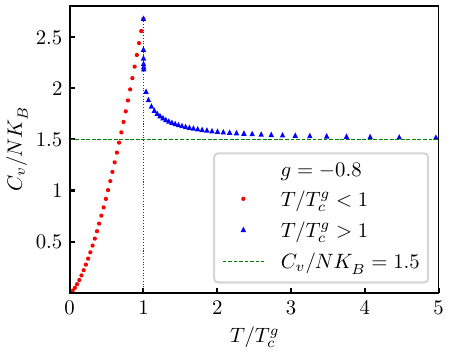}
		\caption{}
		\label{fig:c}
	\end{subfigure}
	\hfill
	\begin{subfigure}[b]{0.49\columnwidth}
		\centering
		\includegraphics[width=\textwidth]{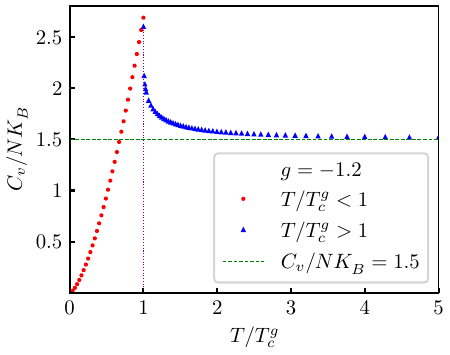}
		\caption{}
		\label{fig:d}
	\end{subfigure}
	
	\caption{\justifying 
		Heat capacity per particle, $C_V/(N k_B)$, as a function of the reduced temperature $T/T_c^{g}$ for a three-dimensional ideal $g$-inclusion-statistics gas with $\sigma=2$ and different values of the parameter $g = 0, -0.4, -0.8$, and $-1.2$.}
	\label{fig:Cv}
\end{figure}

Figure~\ref{fig:Cv} illustrates the scaled heat capacity per particle, $C_V / N k_B$, versus the reduced temperature $T/T_c^g$. The characteristic cusp (non-differentiability) at $T = T_c^g$ signifies a continuous condensation transition. In the asymptotic high-temperature limit, the heat capacity smoothly recovers the classical equipartition value, $C_V =3N k_B/2$.

\section{CONCLUSIONS}\label{CONCLUSIONS}

We have investigated the thermodynamic geometry of an ideal quantum gas
obeying inclusion statistics, obtained by analytically continuing Haldane fractional exclusion statistics to a negative statistical parameter $g<0$.
In this framework, the grand canonical partition function admits a finite maximum fugacity, and the thermodynamic scalar curvature $R$ is strictly positive for all $g<0$; reflecting effective boson-like attractive statistical interactions, while, as the fugacity approaches its maximum value, $R$ diverges, signaling a phase transition of the
BEC type.

The condensation transition is consistently characterized by three independent criteria; the divergence of the thermodynamic curvature, the saturation of the fugacity at its maximum value, and the non-analytic cusp in the specific heat, all coinciding at the same critical point, which confirms the thermodynamic consistency of the transition. Above the transition, the divergence of $R$ recurs at $T = T_c^g$; inside the condensed phase the fugacity is pinned at its maximum value, the parameter space collapses to one dimension, and the curvature vanishes identically.

A key distinction from the ordinary ideal Bose gas is that the condensation temperature is elevated relative to the bosonic case and increases monotonically as $g$ becomes more negative. Moreover, finite-temperature condensation persists down to the dimensional regime $1/2 < D/\sigma \le 1$, which is inaccessible to standard bosons, while for $D/\sigma \le 1/2$ the transition temperature vanishes; in the marginal case $D/\sigma = 1$, a
closed analytic expression for $T_c^g$ is recovered.

This behavior contrasts sharply with the exclusion branch ($g>0$) studied in Ref. \cite{mirza2010thermodynamic}, where the thermodynamic curvature remains finite throughout the state space, exhibiting no divergences. It is negative at low temperatures or small fugacity, indicating repulsive statistical interactions; for $0<g<0.5$ it crosses over to attractive behavior at high temperatures, while for $0.5<g<1$ the response is entirely fermion-like.
Thus, condensation and a genuine critical point exist only in the inclusion regime, where the divergence of the thermodynamic curvature serves as an independent geometric diagnostic of criticality.

These results provide a detailed and internally consistent thermodynamic picture of inclusion statistics and establish it as a genuine extension of
quantum statistics, with behavioral boundaries that are sharply distinct from those of the exclusion branch.

\section*{Acknowledgments}

The authors would like to express their sincere gratitude to Professor Morteza Nattagh Najafi for the insightful discussions and valuable guidance throughout this research.

\bibliography{ref}

\end{document}